\documentstyle[11pt,paspconf,epsfig]{article}

\begin{document}

\title{The Gas Reservoir for present day Galaxies : Damped Ly$\alpha$ 
Absorption Systems}
\author{J.U. Fynbo and B.Thomsen}
\affil{Institute of Physics and Astronomy, University of \AA rhus, 
DK-8000, Denmark}
\author{P. M\o ller}
\affil{European Southern Observatory, Karl-Schwarzschild-Stra\ss e 2,
W-8046 Garching bei M\"unchen, Germany}

\begin{abstract}
  We present results from an ongoing search for galaxy counterparts of a
subgroup of Quasar Absorption Line Systems called Damped Ly$\alpha$
Absorbers (DLAs). DLAs have several characteristics that make them
essential in the process of understanding how galaxies formed in the
early universe and evolved to the galaxies we see today in the local
universe.
   Finally we compare DLAs with recent findings of a
population of starforming galaxies at high redshifts, so called
Lyman-break galaxies.
\end{abstract}


\section{Introduction}
   Damped Ly$\alpha$ Absorbers are the objects with highest HI column density  
of QSO absorption line systems. QSO absorption line systems are intergalactic 
material or in rare cases even galaxies that lie along the line of sight to 
background QSOs. In 
the spectrum of the background QSOs QSO absorption line systems manifest 
themselves primarily in
thousands of Ly$\alpha$ absorption lines on the blue side of the QSO Ly$\alpha$
emission line - the so called Ly$\alpha$ forest. In a simplified picture each 
Ly$\alpha$ absorption line represents an intersecting intergalactic cloud. 
Ly$\alpha$ 
absorption lines at a wavelength close to the QSO Ly$\alpha$ emission line are 
caused by clouds near the QSO in physical space whereas Ly$\alpha$ 
absorption lines further towards the blue are less redshifted and hence caused 
by clouds closer to us along the line of sight (for a recent review see
Rauch, 1998).  

   Damped Ly$\alpha$ Absorbers (DLAs) are QSO absorption line systems are
 causing damped Ly$\alpha$ 
absorption. To do that DLAs have neutral hydrogen column densities larger than 
$2\times10^{20} cm^{-2}$, which is comparable to the column density of baryons
in normal disk galaxies at the present epoch. A very important result recently 
found is that most of the baryons that reside in stars in galaxies today, at 
high redshift were in cold gas in DLAs (Wolfe et al. 1995). In other
words, DLAs constitute the gas reservoir out of which present day galaxies
formed. 

   Since DLAs are objects found by the absorption they cause, much information 
has been collected about metallicity and dust content through the study of 
line strengths of metal lines associated with the DLAs allthough the 
interpretation of the data is still subject to debate (Lu et al., 1996, 
Kulkarni et al., 1997). Little, however, is known about 
the sizes and morphologies of the objects. One way to obtain this information 
is to detect emission from them.

   From an observational point of view the main problems in studying emission 
from DLAs are {\it (i)} that they are very faint and {\it (ii)} the presence of
a much brighter QSO at a distance of only 0-3 arcsec  on the sky. At a redshift
of $z=2$ DLAs produce regions of 15-25\AA \ (the width depending on the HI 
column density) of saturated absorption in the spectrum of the background QSOs. 
Hence imaging in a narrow filter with a width corresponding to the width of the
damped absorption line will circumvent problem {\it (ii)}. If the DLA is a 
Ly$\alpha$ {\it{emitter}} it will be relatively easy to detect against the 
modest sky background in the narrow band filter which circumvents problem 
{\it (i)}. Narrow band imaging of DLAs have been pursued in more than a decade 
(e.g. Lowenthal et al., 1995), but only recently with success. The DLA at 
$z = 2.81$ towards PKS0528-250 (M\o ller and Warren, 1993, 1998, Warren and 
M\o ller, 1996) and the DLA at $z = 1.934$ towards Q0151+048A 
(M\o ller, Warren and Fynbo, 1998, Fynbo, M\o ller and Warren, 1998,
1999) have been detected using the narrow filter technique. 
In the case of the DLA towards Q0151+048A we detected extended Ly$\alpha$ 
emission, which allowed us to obtain the rotation curve of the galaxy (M\o
ller, Fynbo and Warren in prep.).

   In this paper we report on results from a new narrow band project aimed at
the DLA at $z = 1.943$ towards PKS1157+014, and compare results for the DLAs 
with a sample of high redshift galaxies that are selected in a completely 
independent way - the Lyman-break galaxies.  

\section{Observations}

   PKS1157+014 was observed with the 2.56m Nordic Optical Telescope (NOT) 
March 28 - 31 1998. Two of the nights were lost to bad weather. We obtained 
a total integration time of 10 hours in narrow band, and 4000 sec in both I 
and U. The seeing ranged from 0.6 arcsec in I to 0.9 arcsec in the narrow band. 
Due to the two nights lost to bad weather we didn't reach the flux-limits we 
aimed at. With the data obtained we reach a 5$\sigma$ flux limit of 
$7.5\times10^{-17}$ erg s$^{-1}$ cm$^{-2}$ in the narrow band and 5$\sigma$ 
limiting magnitudes of 25.9 and 25.3 in I(AB) and U(AB) respectively.

\section{The field of PKS1157+014}

\begin{figure}[h]
 \begin{center}
 \epsfig{file=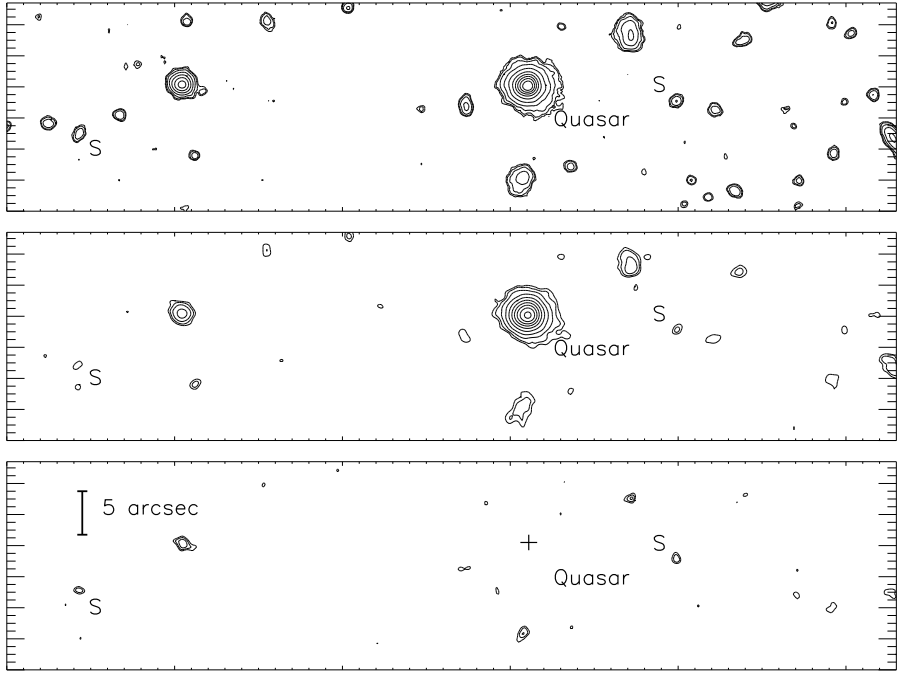,width=12cm}
 \end{center}
 \caption{{\it Top panel :} $96\times24$ arcsec$^2$ surrounding PKS1157+014 
 from 4000sec of integration in the I-band. Marked are the quasar and two 
 candidate  emission line galaxies (S).  North is to the left and east is down.
{\it Middle and lower panels :} The same field as above, but from 4000sec of 
integration in the U-band and 10 hours of integration in the narrow filter.}
\label{field}
\end{figure}

   Fig.~\ref{field} shows $96\times24$ arcsec$^2$ surrounding PKS1157+014 from
the combined I-band, U-band and narrow filter frames. As seen the quasar is
not present in the narrow band frame due to the strong absorption line. We do 
not see any significant Ly$\alpha$ emission at or near the position of the 
quasar from the DLA. We have obtained two more nights on NOT in March 1999, 
which will allow us to reach a 5$\sigma$ detection limit of $5\times10^
{-17} erg s^{-1}$ 
cm$^{-2}$, which is sufficiently deep to detect the DLAs we have seen in 
earlier projects. However, we do detect two candidate emission line galaxies 
marked by 'S' at signal-to-noise levels between 4 and 5 in the combined 
narrow-band frame. 
These very blue and compact emission line galaxies are very similar to the 
emission line galaxies associated to the DLAs seen in the fields of PKS0528-250
and Q0151+048. In both DLA-fields we have studied so far with narrow
band imaging we have found one or more galaxy at the redshift of the DLA,
indicating that also at 
high redshift galaxies were members of groups. It is interesting to note how 
the galaxies seem to be aligned. This is seen in most high redshift groups of 
Ly$\alpha$ emitting galaxies (see Fig. 6 in M\o ller and Warren, 1998).
This trend is in agreement with N-body simulations of hierarchical structure 
formation were galaxies predominantly form along filaments (e.g. Evrard et 
al., 1994). 

\section{Are Lyman-break galaxies and DLAs the same objects?}

   In the last few years hundreds of high redshift galaxies have been found 
using a technique completely independent of QSO absorption lines. This 
technique is based on the fact that young, starforming galaxies will have 
a strong spectral break at the lyman limit, which at high redshift is 
redshifted into the optical window (see Dickinson, 1998, for a recent 
review). Galaxies found using this method are refered to as Lyman-break 
galaxies (LBGs). LBGs need to be bright enough for
spectroscopical confirmation of their high redshift so they are typically
brighter
than R(AB)=26. Since DLAs and LBGs are selected completely independently 
from the population of progenitor galaxies it is very interesting to compare 
the recent results for the LBGs with results from studies of DLAs. 
Assuming that DLAs arise in gaseous discs associated with LBGs one way to 
perform this comparison is to calculate how faint we need to integrate down 
the extrapolation
of the luminosity function of LBGs in order to explain the observed 
probability for a QSO line of sight to cross a DLA. 

    Results of this calculation are presented in Fynbo et al., 1999, and 
summarised here. At $z = 3$ we find that 70-90\% of DLA galaxy counterparts 
are fainter than R(AB)=26, which is the current limit for spectroscopic
confirmation of LBG candidates.
Since DLAs contain close to all the gas that make up present day galaxies 
we conclude that 
the progenitors of a typical present day galaxy at $z = 3$ were small and 
faint and that the LBGs only constitute the tip of the iceberg of high 
redshift galaxies in terms of locating the reservoir of cold gas out of
which present day galaxies formed. This is also consistent with the results 
from semi-analytical modeling of galaxy formation in which LBGs form in very 
rare high overdensity regions and are the progenitors of present day bright 
cluster galaxies (e.g. Baugh et al., 1998). Hence when we wish to study 
properties such as metallicity, dust content and star formation for the 
population of progenitor galaxies as a whole, the DLAs are more likely
representative than LBGs.

\end{document}